\documentclass[draft]{agujournal2019}
\usepackage{url} 
\usepackage{lineno}
\usepackage{soul}
\usepackage{pdflscape}
\usepackage{multirow}
\usepackage{hhline}
\usepackage{lscape}
\usepackage{longtable}
\usepackage{caption}
\usepackage{comment}
\usepackage{setspace}
\usepackage{tablefootnote}
\usepackage{amsmath}
\usepackage{footnotehyper}
\usepackage{dcolumn}
\usepackage{siunitx}
\draftfalse
\journalname{JGR: Planets}

\begin{document}
%
%


\title{Nonthermal hydrogen loss at Mars: Contributions of photochemical mechanisms to escape and identification of key processes}

%
%




\authors{Bethan S. Gregory\affil{1}, Michael S. Chaffin\affil{1}, Rodney D. Elliott\affil{1}, Justin Deighan\affil{1}, Hannes Gr\"oller\affil{2}, and Eryn Cangi\affil{1}}


\affiliation{1}{Laboratory for Atmospheric and Space Physics, University of Colorado Boulder, Boulder, Colorado, USA}
\affiliation{2}{Lunar and Planetary Laboratory, University of Arizona, Tucson, Arizona, USA}




\correspondingauthor{Bethan S. Gregory}{bethan.gregory@colorado.edu}





\begin{keypoints}
\item We quantify nonthermal hydrogen escape from 47 sources, many for the first time, using Monte Carlo model-derived escape probability profiles
\item Total nonthermal escape from these sources is 27--39\% of the thermal escape flux of 8.7 \texttimes 10$^7$ -- 1.1 \texttimes 10$^8$ cm$^{-2}$ s$^{-1}$
\item HCO$^+$ dissociative recombination is the biggest contributor to photochemical H escape at Mars, and likely at Venus and some exoplanets

\end{keypoints}

\noindent This is the accepted manuscript. The published version can be found at:
Gregory, B.S., Chaffin, M.S., Elliott, R.D., Deighan, J., Gröller, H. and Cangi, E., 2023. Nonthermal hydrogen loss at Mars: Contributions of photochemical mechanisms to escape and identification of key processes. Journal of Geophysical Research: Planets, p.e2023JE007802. https://doi.org/10.1029/2023JE007802.

%
%

%
%


\begin{abstract}
Hydrogen loss to space is a key control on the evolution of the Martian atmosphere and the desiccation of the red planet. Thermal escape is thought to be the dominant loss process, but both forward modeling studies and remote sensing observations have indicated the presence of a second, higher-temperature ``nonthermal" or ``hot" hydrogen component, some fraction of which also escapes. Exothermic reactions and charge/momentum exchange processes produce hydrogen atoms with energy above the escape energy, but H loss via many of these mechanisms has never been studied, and the relative importance of thermal and nonthermal escape at Mars remains uncertain. Here we estimate hydrogen escape fluxes via 47 mechanisms, using newly-developed escape probability profiles. We find that HCO$^+$ dissociative recombination is the most important of the mechanisms, accounting for 30--50\% of the nonthermal escape. The reaction CO$_2^+$ + H$_2$ is also important, producing roughly as much escaping H as momentum exchange between hot O and H. Total nonthermal escape from the mechanisms considered amounts to 39\% (27\%) of thermal escape, for low (high) solar activity. Our escape probability profiles are applicable to any thermospheric hot H production mechanism and can be used to explore seasonal and longer-term variations, allowing for a deeper understanding of desiccation drivers over various timescales. We highlight the most important mechanisms and suggest that some may be important at Venus, where nonthermal escape dominates and much of the literature centers on charge exchange reactions, which do not result in significant escape in this study.
\end{abstract}

\section*{Plain Language Summary}
The climate of Mars has become drier over time, partially due to hydrogen and oxygen atoms escaping from the planet's upper atmosphere. This can happen when the atoms have sufficient energy to overcome Mars' gravitational pull. Some fraction of escaping hydrogen---known as ``hot'' hydrogen---gains this energy through chemical processes, but the extent to which this contributes to the total hydrogen escape rate remains uncertain. Here we estimate the rate of hydrogen escape from 47 different chemical mechanisms, by considering the altitude above the surface at which they are produced and the energy that they might be given when they are formed. We find the mechanisms producing the most escaping hydrogen and estimate how quickly hot hydrogen is being lost to space. The most important mechanism, of those we consider, is HCO$^+$ + e$^{\text{-}}$ $\rightarrow$ CO + H, which contributes up to half of the escaping hot H. We speculate that this is also important at Venus.

%
%

%


%
%
%
%

\section{Introduction}
 The Martian environment has changed markedly over its four-and-a-half billion year history \cite<e.g.>{Carr2010,Jakosky2001}. Evidence for a high initial surface water inventory comes from geomorphology and mineralogy studies, including observations of delta deposits, fluvial features, and crater lacustrine sediments \cite<e.g.>{Bibring2006,Ehlmann2014,Goudge2017,Grotzinger2015,Squyres2004}. Since then, the escape of H and O to space has driven considerable desiccation; deuterium to hydrogen ratios indicate that at least 85\% of the water has been lost \cite{Owen1988,Villanueva2015}. Given the present-day water inventory at Mars, measured D/H ratios require a water sink of tens to hundreds of meters GEL (global equivalent layer, i.e., the depth of the reservoir if spread uniformly across Mars' surface) via escape. For example, \citeA{Alsaeed2019} predict loss of 20--220 m GEL over the last 3.3 billion years, and Cangi et al. (2023, this issue) predict loss of 147--158 m GEL over Mars' history. Extrapolation of modern hydrogen escape rates back in time cannot account for water loss of this magnitude, and one potential solution is that escape was much higher early in Mars' history (Cangi et al., 2020; Carr \& Head, 2019; Cravens et al., 2017; Jakosky et al., 2018; Lammer et al., 2003; 2006a, Lillis et al., 2017). Understanding the rates of and controls on modern volatile escape improves understanding of its variation in the past. This in turn can aid estimation of the integrated water sink from escape and the initial water inventory at formation, which have important astrobiological implications.
\nocite{Alsaeed2019,Cangi2020,Carr2019,Cravens2017,Jakosky2018,Lammer2003,Lammer2006a,Lillis2017}

Thermal (Jeans) escape, in which a proportion of particles in the high-energy tail of a Maxwell-Boltzmann velocity distribution have energies greater than escape energy, is the main loss process for neutral hydrogen at Mars. Present-day derived thermal escape rates vary seasonally \cite<between 0.1--50\texttimes 10$^8$ cm$^{-2}$ s$^{-1}$;>{Bhattacharyya2015, Bhattacharyya2017,Chaffin2014,Chaufray2015,Chaufray2021,Clarke2014, Halekas2017,Jakosky2018,Mayyasi2022} and over solar cycles \cite<e.g.,>{Chaufray2015,Fox2015,Krasnopolsky2019}. Averaged over seasonal and solar activity timescales, hydrogen fluxes of up to 2.4 \texttimes 10$^8$ cm$^{-2}$ s$^{-1}$, set to balance O escape in a 2:1 ratio, can be obtained from equilibrium photochemical models \cite{Nair1994}. Neutral species can alternatively gain sufficient energy to escape when they are produced by exothermic photochemical reactions and charge or momentum exchange processes with higher-temperature particles; the resulting energetic atoms are ``nonthermal'' or ``hot.'' Nonthermal escape, primarily via O$_2^+$ dissociative recombination, dominates oxygen escape at Mars (Fox \& Hać, 2009; Leblanc et al., 2017; Lillis et al., 2017; Luhmann, 1997; McElroy, 1972; Nagy \& Cravens, 1988). Photochemical processes also dominate carbon escape, primarily via CO$_2$ photodissociation (Lo et al., 2021, 2022; Thomas et al., 2023). However, for H, the importance of hot escape relative to thermal escape has yet to be fully determined.
\nocite{Fox2009,Leblanc2017,Lillis2017,Luhmann1997,McElroy1972,Nagy1988}
\nocite{Lo2021,Lo2022,Thomas2023}

The uncertainty surrounding the contribution of hot H to total escape is partly due to the lack of direct observations of hot H at Mars. Unlike at Venus \cite{Anderson1976,Kumar1974}, its presence has remained ambiguous, as greater hydrogen scale heights and larger thermal population densities mean that contributions to observed brightnesses from each of the two populations are not so easily distinguished. Estimates for hot H densities and escape rates therefore come from two sources: (1) retrievals from Lyman alpha brightness observations made by remote sensing instruments, which must be processed by radiative transfer models, and (2) model calculations predicting hot H densities and escape rates from known processes and physical parameters. The presence of hot H in Mars' corona has been indicated by interpretations of brightness observations, some of which are better explained by models incorporating both thermal and nonthermal hydrogen components with different temperatures \cite{Bhattacharyya2015,Bhattacharyya2017Icarus,Bhattacharyya2017,Chaffin2018,Chaufray2008,Galli2006,Lichtenegger2004}.

Despite the observational evidence for hot H at Mars, its densities are difficult to constrain. Hot H exobase densities of up to half of the thermal density have been retrieved by some authors (Bhattacharyya et al., 2015 (8--30\%); Chaufray et al., 2008 (6--20\%); Galli et al., 2006 (50\%)), whereas other retrievals predict a low-density hot population of less than 1\% of the total at the exobase \cite{Chaffin2018}. Meanwhile, some predicted total hydrogen escape rates double with the inclusion of a hot population in certain seasonal conditions \cite{Bhattacharyya2015,Bhattacharyya2017Icarus,Bhattacharyya2017}. One of the sources of uncertainty in such density and escape rate estimates is that they are generally derived from observations by assuming a Maxwellian distribution; the escape rates are sensitive to the chosen temperature for the distribution, but this can be difficult to constrain.

In general, modeling studies have predicted lower hot H densities and escape rates than retrievals. Several predictions of densities and escape from photochemical sources are less than 5\% of thermal densities and escape \cite{Krasnopolsky2010,Lichtenegger2006,Nagy1990}. Some of the mechanisms previously considered include charge exchange of hydrogen ions with neutral O or H atoms (Lichtenegger et al., 2006; Nagy et al., 1990):\\
\begin{equation}
    \mathrm{H_{hot}^+ + H \rightarrow H_{hot} + H^+}
\end{equation}
\begin{equation}
    \mathrm{H_{hot}^+} + \mathrm{O} \rightarrow \mathrm{H_{hot}} + \mathrm{O^+}
\end{equation}\\

\noindent and the following photochemical reactions \cite{Lichtenegger2006}: \\
\begin{equation}
\mathrm{CO_2^+ + H_2 \rightarrow OCOH^+ + H}
\end{equation}
\begin{equation}
\mathrm{O^+ + H_2 \rightarrow OH^+ + H}
\end{equation}
\begin{equation}
\mathrm{OCOH^+ + e^{\text{-}} \rightarrow CO_2 + H}
\end{equation}
\begin{equation}
\mathrm{OH^+ + e^{\text{-}} \rightarrow O + H}.
\end{equation}

However, some models have highlighted two mechanisms which individually account for considerable escape. \citeA{Shematovich2013} predicted nonthermal escape equal to 5\% of their assumed thermal escape flux (1.1 \texttimes 10$^8$ cm$^{-2}$ s$^{-1}$) via momentum exchange with hot oxygen:\\
\begin{equation}
    \mathrm{O_{hot} + H \rightarrow O + H_{hot}}.
\end{equation}
\noindent  An additional reaction recently explored for the first time for Mars is HCO$^+$ dissociative recombination:\\
\begin{equation}
\mathrm{HCO^+ + e^{\text{-}} \rightarrow CO + H},
\end{equation}
for which global escape rates of up to 14\% of the same assumed thermal escape rate were predicted (Gregory et al., 2023a). The latter results highlight the possibility that other mechanisms, which have not been studied before, might also contribute significantly to hot H escape. For example, Fox's (2015) photochemical ionosphere model includes dozens of exothermic reactions with H as a product and excess energies above the escape energy (0.13 eV at an altitude of 80 km and 0.12 eV at 400 km), but that have not been included in escape models to date.

Here, our aim is to investigate the potential for ion-neutral mechanisms to play a key role in modern and ancient hot H escape at Mars and assess whether the inclusion of these mechanisms in models might reduce the discrepancy with observations. We undertake a systematic study of 47 hot H-producing mechanisms, using new escape probability profiles to estimate escape fluxes of H produced by these processes and highlight the most important. In Section 2, we introduce the mechanisms on which we focus, which are listed in Table 1. In Section 3, we outline the method adopted to estimate escape fluxes from each mechanism, including descriptions of the production rate and escape probability profiles used. The escape flux results are presented in Section 4 and discussed in Section 5.

\section{Sources of nonthermal H at Mars}
The sources of hot H that we consider in this study are shown in Table 1. The mechanisms comprise dissociative recombination and bimolecular reactions from Fox's (2015) photochemical model, as well as charge exchange mechanisms of hot H ions with O or H atoms (R8 and R11). \citeA{Shematovich2013} computed escape of hot H produced by momentum transfer between hot O and H (Equation 7). Because it is complicated to apply the method we outline below to this mechanism and it has already been modeled in detail, we do not model it here, but compare our results to the calculated escape flux (6 \texttimes 10$^6$ cm$^{-2}$ s$^{-1}$). There is also a likely source of nonthermal hydrogen from charge exchange between solar wind protons and atmospheric hydrogen atoms (e.g. Krasnopolsky, 2010; Shematovich, 2021), but we do not include this, instead focusing only on the photochemical reactions and charge exchange processes between atmospherically-derived particles.
\nocite{Shematovich2021}

\begin{singlespace}
\renewcommand{\thetable}{1}
\begin{longtable}{p{5mm}llllll}
\caption{Hot hydrogen-producing mechanisms considered in this study. We have specifically labeled the ``hot'' particles involved in the charge exchange mechanisms (R8 and R11) for clarity, but all mechanisms listed produce hot hydrogen. The excess energy from each reaction (Section 2.1) is shown in Column 3. The initial H kinetic energy assumed for selection of the escape probability profile is shown in Column 4. As noted in Section 3, the escape probability profile chosen is that for the energy closest to but smaller than the reaction's excess energy. Consideration of the details of the exact kinetic energies imparted to the hydrogen atoms by each process (e.g., incorporation of the internal degrees of freedom of the products) is beyond the scope of this work. The 47 mechanisms are listed and numbered in order of largest escape flux for the low solar activity case (for results, see Section 4).}\\
\hline
&&&\\
&   &        & Assumed initial energy    \\
&   & Actual excess        & for escape probability          \\
\# & Source & energy [eV] & profiles [eV] \\
&&\\
\hline
\endfirsthead

\multicolumn{6}{c}%
{{\bfseries Table \thetable\ continued from previous page}} \\
\hline
&&&\\
&   &        & Assumed initial energy    \\
&   & Actual excess        & for escape probability          \\
\# & Source & energy [eV] & profiles [eV] \\
&&\\
\hline
\endhead
&&\\
R1 & HCO$^+$ + e$^{\text{-}}$ $\rightarrow$ CO + H                 & 7.31 & 5    \\
R2 & CO$_2^+$ + H$_2$ $\rightarrow$ OCOH$^+$ + H            & 1.35 & 1    \\ 
R3 & O$^+$($^4$S) + H$_2$ $\rightarrow$ OH$^+$ + H                 & 0.54 & 0.5   \\ 
R4 & OH$^+$ + O $\rightarrow$ O$_2^+$ + H                 & 1.65 & 1    \\
R5 & OCOH$^+$ + e$^{\text{-}}$ $\rightarrow$ O + CO + H            & 2.42 & 1    \\
R6 & N$_2^+$ + H$_2$ $\rightarrow$ N$_2$H$^+$ + H             & 2.61 & 1     \\ 
R7 & O$^+$($^2$D) + H$_2$ $\rightarrow$ OH$^+$ + H                 & 3.86 & 1   \\ 
R8 & O + H$_{\textrm{hot}}^+$ $\rightarrow$ O$^+$ + H$_{\textrm{hot}}$        & $\dag$ & 0.2  \\
R9 & H$_2^+$ + O $\rightarrow$ OH$^+$ + H                & 2.35 & 1    \\
R10 & CO$^+$ + H$_2$ $\rightarrow$ HCO$^+$ + H              & 2.04 & 1   \\ 
R11 & H + H$_{\textrm{hot}}^+$ $\rightarrow$ H$^+$ + H$_{\textrm{hot}}$        & $\dag$ & 0.2  \\
R12 & CO$^+$ + H$_2$ $\rightarrow$ HOC$^+$ + H              & 0.61 & 0.5 \\ 
R13 & H$_2^+$ + CO$_2$ $\rightarrow$ OCOH$^+$ + H           & 3.00 & 1    \\ 
R14 & OCOH$^+$ + e$^{\text{-}}$ $\rightarrow$ CO$_2$ + H           & 7.95 & 5     \\
R15 & OH$^+$ + H$_2$ $\rightarrow$ H$_2$O$^+$ + H             & 1.02 & 1     \\ 
R16 & H$_2^+$ + N$_2$ $\rightarrow$ N$_2$H$^+$ + H            & 2.43 & 1   \\ 
R17 & Ar$^+$ + H$_2$ $\rightarrow$ ArH$^+$ + H              & 1.60 & 1    \\
R18 & N$^+$ + H$_2$ $\rightarrow$ NH$^+$ + H                & 3.09 & 1     \\ 
R19 & N$_2$H$^+$ + e$^{\text{-}}$ $\rightarrow$ N$_2$ + H              & 8.47 & 5   \\
R20 & CH$^+$ + O $\rightarrow$ CO$^+$ + H                 & 4.27 & 1    \\
R21 & H$_2^+$ + H$_2$ $\rightarrow$ H$_3^+$ + H             & 1.71 & 1     \\ 
R22 & O$^+$($^2$P) + H$_2$ $\rightarrow$ OH$^+$ + H                 & 5.56 & 5   \\ 
R23 & OH$^+$ + N $\rightarrow$ NO$^+$ + H                 & 5.84 & 5    \\ 
R24 & H$_2^+$ + CO $\rightarrow$ HCO$^+$ + H              & 3.47 & 1    \\ 
R25 & H$_2^+$ + CO $\rightarrow$ HOC$^+$ + H              & 2.05 & 1     \\
R26 & H$_2^+$ + Ar $\rightarrow$ ArH$^+$ + H              & 1.27 & 1     \\
R27 & OH$^+$ + e$^{\text{-}}$ $\rightarrow$ O($^1$D) + H             & 6.63 & 5     \\
R28 & H$_3^+$ + O $\rightarrow$ H$_2$O$^+$ + H              & 1.66 & 1    \\
R29 & H$_2^+$ + N $\rightarrow$ NH$^+$ + H                & 3.98 & 1     \\
R30 & H$_2^+$ + O$_2$ $\rightarrow$ HO$_2^+$ + H            & 1.67 & 1    \\ 
R31 & H$_2^+$ + e$^{\text{-}}$ $\rightarrow$ H + H                 & 10.95 & 10   \\
R32 & O$^+$($^2$P) + H$_2$ $\rightarrow$ O + H$^+$ + H                 & 0.52 & 0.5   \\ 
R33 & CH$^+$ + e$^{\text{-}}$ $\rightarrow$ C($^1$D) + H             & 5.92 & 5     \\
R34 & CH$^+$ + H$_2$ $\rightarrow$ CH$_2^+$ + H             & 0.18 & 0.2   \\
R35 & H$_2^+$ + C $\rightarrow$ CH$^+$ + H                & 3.78 & 1   \\
R36 & CH$^+$ + e$^{\text{-}}$ $\rightarrow$ C($^1$S) + H             & 4.50 & 1     \\
R37 & H$_3^+$ + e$^{\text{-}}$ $\rightarrow$ H + H + H             & 4.76 & 1    \\
R38 & HNO$^+$ + e$^{\text{-}}$ $\rightarrow$ NO + H                & 7.97 & 5    \\
R39 & CH$^+$ + N $\rightarrow$ CN$^+$ + H                 & 0.80 & 0.5    \\
R40 & CH$^+$ + C $\rightarrow$ C$_2^+$ + H                & 1.27 & 1     \\
R41 & H$_3^+$ + e$^{\text{-}}$ $\rightarrow$ H$_2$ + H               & 9.23 & 5    \\
R42 & HO$_2^+$ + e$^{\text{-}}$ $\rightarrow$ O$_2$ + H              & 9.25 & 5     \\
R43 & HO$_2^+$ + e$^{\text{-}}$ $\rightarrow$ O + O + H            & 4.13 & 1    \\
R44 & ArH$^+$ + e$^{\text{-}}$ $\rightarrow$ Ar + H                & 9.68 & 5     \\
R45 & He$^+$ + H$_2$ $\rightarrow$ H$^+$ + He + H & 6.51 & 5 \\
R46 & HNO$^+$ + O $\rightarrow$ NO$_2^+$ + H              & 1.33 & 1     \\
R47 & HO$_2^+$ + N $\rightarrow$ NO$_2^+$ + H             & 4.00 & 1     \\
&&\\
\hline
%
\multicolumn{4}{l}{$^{\dag}$ See Section 2.1 and Supplementary Information for a description of the}\\
\multicolumn{4}{l}{energies of H atoms produced by the two charge exchange mechanisms.}
\end{longtable}

\end{singlespace}

\subsection{Excess energies of source mechanisms}
In order for source photochemical reactions to produce hot hydrogen, they must be exothermic and release excess energy, some of which can be imparted to the hydrogen atom as translational kinetic energy. The excess energy, $E_{\textrm{excess}}$, of a reaction is calculated using the enthalpies of formation ($\Delta H^0_f$) of the reactants (A and B) and products (C and D):

\begin{equation}
    E_{\textrm{excess}} = \Delta H^0_f[A] + \Delta H^0_f[B] - (\Delta H^0_f[C] + \Delta H^0_f[D])
\end{equation}

We take enthalpy of formation values from the UMIST Astrochemistry Database (McElroy et al., 2013) and the Argonne National Laboratory Active Thermochemical Tables \cite{Ruscic2005,Ruscic2021} and assume that the enthalpy of formation for electrons is zero. For R33 and R36 (CH$^+$ dissociative recombination reactions), which produce excited-state carbon atoms, we use the values for kinetic energy release from \citeA{Amitay1996}. Assuming, except where noted explicitly, that all products are in the ground state, the calculated excess energies for all mechanisms are shown in Column 3 of Table 1, with the largest excess energy, 10.95 eV, from H$_2^+$ dissociative recombination (R31). There is one mechanism from Fox's (2015) compilation for which we do not estimate escape fluxes (C$^+$ + H$_2$ $\rightarrow$ CH$^+$ + H; not shown in Table 1), as it is endothermic at room temperature and therefore unlikely to constitute a significant source of hot hydrogen in Mars' atmosphere.

\nocite{McElroy2013}

Hydrogen atoms produced by the charge exchange reactions (R8 and R11), in which an H$^+$ is neutralized, have a range of initial energies dependent on the energy of the original ion. For these mechanisms, we assume that the energy distribution of the initial H$^+$ can be described by the Maxwell-Boltzmann velocity distribution at the (altitude-dependent) ion temperature. For resonant charge exchange, we assume that the kinetic energy of the neutralized H is the same as that of the initial ion. For charge exchange between O and H$^+$, we assume that the kinetic energy of the neutralized H is equal to the kinetic energy of the initial ion minus the energy required to ionize the O atom (i.e., the difference in ionization potential between oxygen and hydrogen, \mbox{0.019} eV). The spread in energies of the reactants thus leads to a spread of excess energies of the produced hot H, with mean kinetic energies varying with altitude between 0.02 eV and 0.48 eV, so some  will be produced below the escape energy (0.12--0.13 eV). Note that, since we do not consider momentum exchange between the H ion and the initial neutral species, our energy estimates for these processes are upper limits.

\section{Estimating escape fluxes}
The method we use to calculate escape fluxes is similar to that used by \citeA{Fox2009, Fox2018} and \citeA{Lillis2017} for hot oxygen escape via O$_2^+$ dissociative recombination and is illustrated for two example mechanisms in Figure 1. For each mechanism, we multiply a production rate profile ($R(z)$; panel (a); see Section 3.1) by an escape probability profile ($p(z,E_H)$; panel (b)) to obtain a profile of the production rate of escaping H (panel (c)), where $z$ is the altitude and $E_H$ is the initial kinetic energy of the hydrogen atoms. We integrate under the resulting curve to obtain a hydrogen escape flux, $\phi$:

\begin{equation}
    \hspace{20mm}\phi = \int R(z) p(z,E_H) dz \hspace{10mm} \text{[cm$^{-2}$ s$^{-1}$]}
\end{equation}

The escape probability profiles indicate the likelihood that a hydrogen atom produced at a certain altitude with a certain initial kinetic energy will escape, and they are based on Monte Carlo modeling of energetic hydrogen atoms produced at a series of altitudes at 20 km intervals. We produce five escape probability profiles for different initial hydrogen atom kinetic energies: 0.2 eV, 0.5 eV, 1 eV, 5 eV, and 10 eV. The 0.2 eV and 5 eV curves come from Gregory et al. (2023a). We use the method and model described therein to produce three further escape probability curves (for 0.5 eV, 1 eV, and 10 eV), since many of the reaction excess energies are not close to either 0.2 eV or 5 eV and it is not obvious how best to interpolate between these values, given the structure present in the differential collision cross sections \cite{Zhang2009}. We provide further details of the escape probability profiles in Section 3.2. Since, for simplicity, we do not produce an escape probability profile for each mechanism’s exact excess energy, we use the profile for the energy closest to and lower than the mechanism's excess energy to estimate the escape flux. For example, the excess energy from HCO$^+$ dissociative recombination (R1) is 7.31 eV, so we choose the 5 eV escape probability profile to represent the likelihood of escape of the H atoms produced by this mechanism. We round down rather than up to energies because the excess energy is an upper limit for the kinetic energy imparted to the H. The mechanism's actual excess energy and the excess energy assumed in the choice of escape probability profile are listed in Columns 3 and 4 of Table 1, respectively.

We stress that this is a simple method for quickly determining a first-order estimate of escape fluxes from different mechanisms. For this reason, a detailed analysis of the likely velocities of the H atoms is beyond the scope of this work. Specifically, we neglect the following factors: (1) we do not calculate the proportion of the excess energy that goes into the velocity of the H atom as opposed to the velocity of the other product; (2) we do not consider branches of reactions with second products in different electronic, vibrational, and rotational states, instead assuming, except where noted, that all products are in the ground state; and (3) we ignore energy imparted to the hydrogen atom from the initial velocities of the reactants (other than for R8 and R11, see below), as this is at most $\sim$0.5 eV. We discuss the effects of these assumptions in Section 5. Despite these simplifications, our work represents the most comprehensive survey of hot H production mechanisms yet undertaken.

\begin{figure}[!ht]
    \centering
    \includegraphics[scale=0.4]{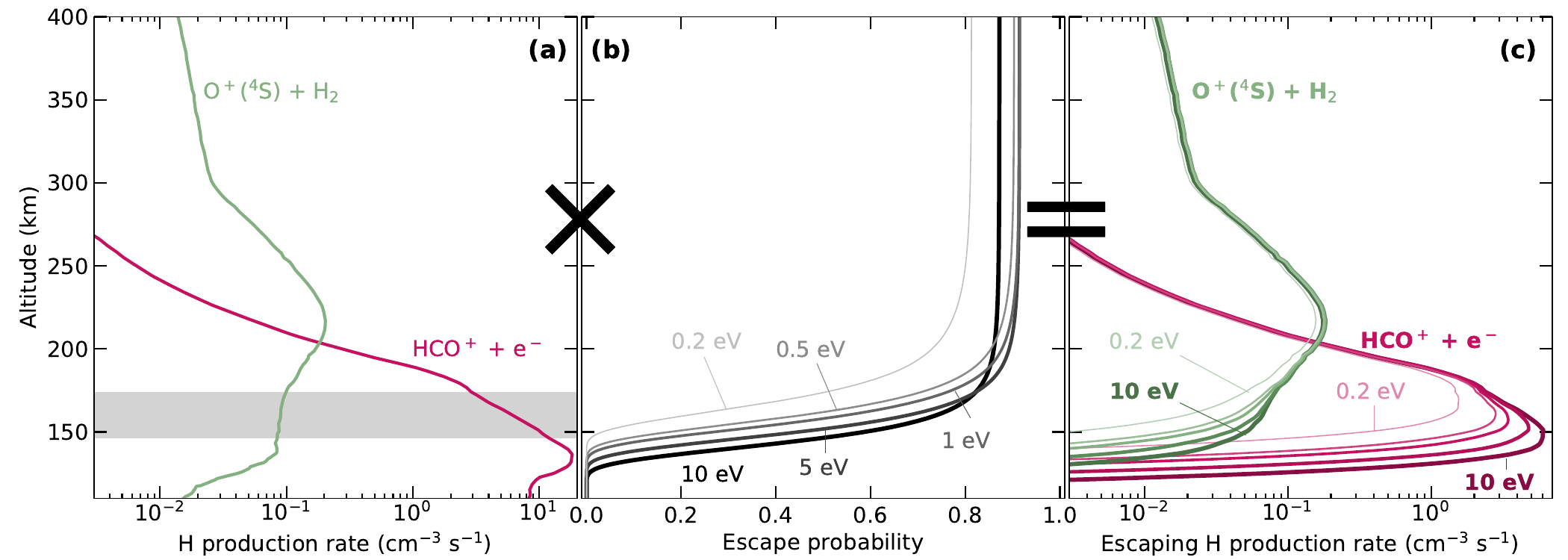}
    \caption{Method for estimating escape fluxes. Production rate profiles (a) are multiplied by escape probability profiles (b) to obtain profiles of the production rate of escaping hot H (c). Two example hot H-producing mechanisms are shown: HCO$^+$ dissociative recombination (pink; R1 in Table 1) and O$^+$($^4$S) + H$_2$ (green; R3). Escape probability profiles for five initial H atom kinetic energies are presented, indicating the likelihood that a hydrogen atom produced at a certain altitude with a certain energy will escape. The escaping H production rate profiles are shown for each of the two example mechanisms and for all five initial energies, though in reality the true profile will be controlled by the specific energy spectrum of the production mechanism, likely falling between the curves shown. For reference, the gray shading in panel (a) indicates the altitudes above which 50\% of the produced particles escape, for particles with initial energies between 0.2 eV and 10 eV, according to the curves in panel (b). In panels (b) and (c), paler colors indicate smaller initial energies. The examples shown are for low solar activity conditions.}
    \label{fig:1_method}
\end{figure} 

For the two charge exchange mechanisms, R8 and R11, we assume that the energies of the product hydrogen will follow a Maxwellian distribution, due to the initial H$^+$ distribution at the ion temperature, $T_i$. Some of the particles will be produced below the escape energy, so we integrate over the energy distribution to calculate a fraction, $q$, of particles produced above the escape energy. For further details, see the Supplementary Material and Figure S1. The escape flux for R8 and R11 is then:\\
\begin{equation}
    \hspace{28mm}\phi = \int R(z) p(z,E_H) q(T_i(z)) dz \hspace{10mm}\textrm{[cm$^{-2}$ s$^{-1}$] (R8 and R11 only)}
\end{equation}
\noindent which discounts H produced below the escape energy in the escape estimate calculation.

\subsection{Production rates of source mechanisms}
Hot H production rates for each mechanism are calculated using rate coefficients from Fox (2015), Rodriguez et al. (1984), and the UMIST Astrochemistry database (McElroy et al., 2013; see Table S1) and temperature profiles from Fox (2015). Density profiles (Figure S2) come from Fox's (2015) ``eroded'' photochemical model, except for the O$^+$($^2$D) and O$^+$($^2$P) densities, which come from \citeA{Fox1996}. Densities of ground-state O$^+$($^4$S) are then assumed to be those of O$^+$($^2$D) and O$^+$($^2$P) subtracted from those of all O$^+$ states combined (from Fox, 2015). The lower and upper limits of the input profiles mean that production is simulated at altitudes of 80--400 km. We extrapolate Fox's (2015) temperature profiles by assuming the 100 km value for lower altitudes. The hot H production rate at altitude $z$, $R(z)$, is then:\\
\begin{equation}
    R(z) = x\hspace{0.8mm}k(z) \hspace{0.8mm} n_1(z) \hspace{0.8mm} n_2(z)
\end{equation}
where $x$ is the number of hydrogen atoms produced by the mechanism, $k$ is the rate coefficient, often dependent on temperature and therefore altitude, and $n_1$ and $n_2$ are the reactant densities.

\nocite{Fox2015,Rodriguez1984,McElroy2013}

We calculate production for both low and high solar activity conditions, defined by Fox (2015), who chose a 60\textdegree \hspace{0.4mm} solar zenith angle (equal to the area-weighted mean dayside solar zenith angle), a Mars--Sun distance of 1.524 AU, and F$_{10.7}$ values of 68 and 214, respectively. Our variation of solar activity is entirely through using the input density and temperature profiles for each case from this photochemical model.

\subsection{Escape probability profiles}

In order to produce escape probability profiles, we fit the Monte Carlo model-generated escape probability output (see above) to curves of the form $p = Ae^{-bN\sigma}$, dependent on the background column density overhead, $N$, and the total cross section at the chosen energy, $\sigma$. The constants $A$ and $b$ represent the escape probability with no overhead atmosphere, and a transparency coefficient, respectively. The escape probability, $p$, is then:\\
\begin{gather}
\begin{flalign}
   &p = 0.815e^{-0.127N\sigma} \hspace{24mm}
   \end{flalign}
\end{gather}
\noindent for H atom initial kinetic energies of 0.2 eV, \\
\begin{gather}
\begin{flalign}
   &p = 0.903e^{-0.069N\sigma} \hspace{4mm} \text{for 0.5 eV,}\\
   &p = 0.915e^{-0.056N\sigma} \hspace{4mm}\text{for 1 eV,}\\
   &p = 0.914e^{-0.039N\sigma} \hspace{4mm}\text{for 5 eV, and} \\
   &p = 0.872e^{-0.023N\sigma} \hspace{4mm}\text{for 10 eV}. 
   \end{flalign}
\end{gather}
\noindent This results in different altitude profiles for the two solar activity cases (Figure 1b; see Figure 3g for high solar activity).

For all energies, more than 82\% of the hot H produced at the upper production altitude (400 km) escapes, despite the initial velocities of the particles being isotropic in direction. This requires 65\% or more of the initially downward-directed particles to be ``reflected" back into space via collisions with the background atmosphere. This reflection must occur before those same collisions cause thermalization of the particles, to which we assign the strict definition of causing a particle's energy to drop below the escape energy, at which point we assume that it will never regain energies sufficient for escape. As seen in Figures 1b/3c and 3f, below 170 km (220 km), for low (high) solar activity, escape probability increases with initial particle energy. Higher initial energies allow these particles to experience more collisions before being thermalized. Higher in the atmosphere, this correlation is not seen. For example, escape probabilities for 10 eV particles above 210 km (275 km) for low (high) solar activity conditions are lower than those for 0.5 eV, 1 eV, and 5 eV. This is because the escape probability is not only affected by the initial energy of the particle, but also by the shape of the total cross section and differential cross sections (Figure S3), which control the likelihood of collision and the resulting scattering angle, respectively. The relative importance of the cross sections and velocities for escape is discussed in Section 5.3. For reference only, the exobase altitude for 5 eV H atoms is at $\sim$177 km and $\sim$204 km for low and high solar activity, respectively, calculated as the point at which:\\
\begin{equation}
    \sigma_{CO_{2}}n_{CO_{2}}H_{CO_{2}} = 1
\end{equation}
where $\sigma_{CO_{2}}$ is the collision cross section between H and CO$_2$ (for which we actually use the O--H total cross section of 3.89 \texttimes 10$^{-15}$ cm$^{2}$), $n_{CO_{2}}$ is the local CO$_2$ number density, and $H_{CO_{2}}$ is the CO$_2$ scale height. However, we note that escape from the Monte Carlo model is independent of the exobase altitude \cite<see, e.g.,>{Fox2009}.

\section{Results: Computed escape flux estimates}
Figure 2 shows the estimated hot H escape fluxes for the five most important mechanisms considered. The dissociative recombination of HCO$^+$ produces the largest escape flux, and mechanisms 2--4 in Table 1---CO$_2^+$ + H$_2$ $\rightarrow$ OCOH$^+$ + H, O$^+$($^4$S) + H$_2$ $\rightarrow$ OH$^+$ + H, and OH$^+$ + O $\rightarrow$ O$_2^+$ + H---are also among the most important processes. The reactions OCOH$^+$ + e$^{\text{-}}$ $\rightarrow$ O + CO + H and N$_2^+$ + H$_2$ $\rightarrow$ N$_{2}$H$^+$ + H are additionally important for low and high solar activity, respectively. For comparison, we show the escape flux of energetic hydrogen produced by momentum exchange (Shematovich, 2013) and assume that this is the same for both low and high solar activity, which puts it within the top three most important mechanisms according to our results. The estimated escape fluxes for the 47 mechanisms, for all five initial particle population energies and both low and high solar activity, are shown in Table S1.

\begin{figure}[!ht]
    \centering
    \includegraphics[scale=0.55]{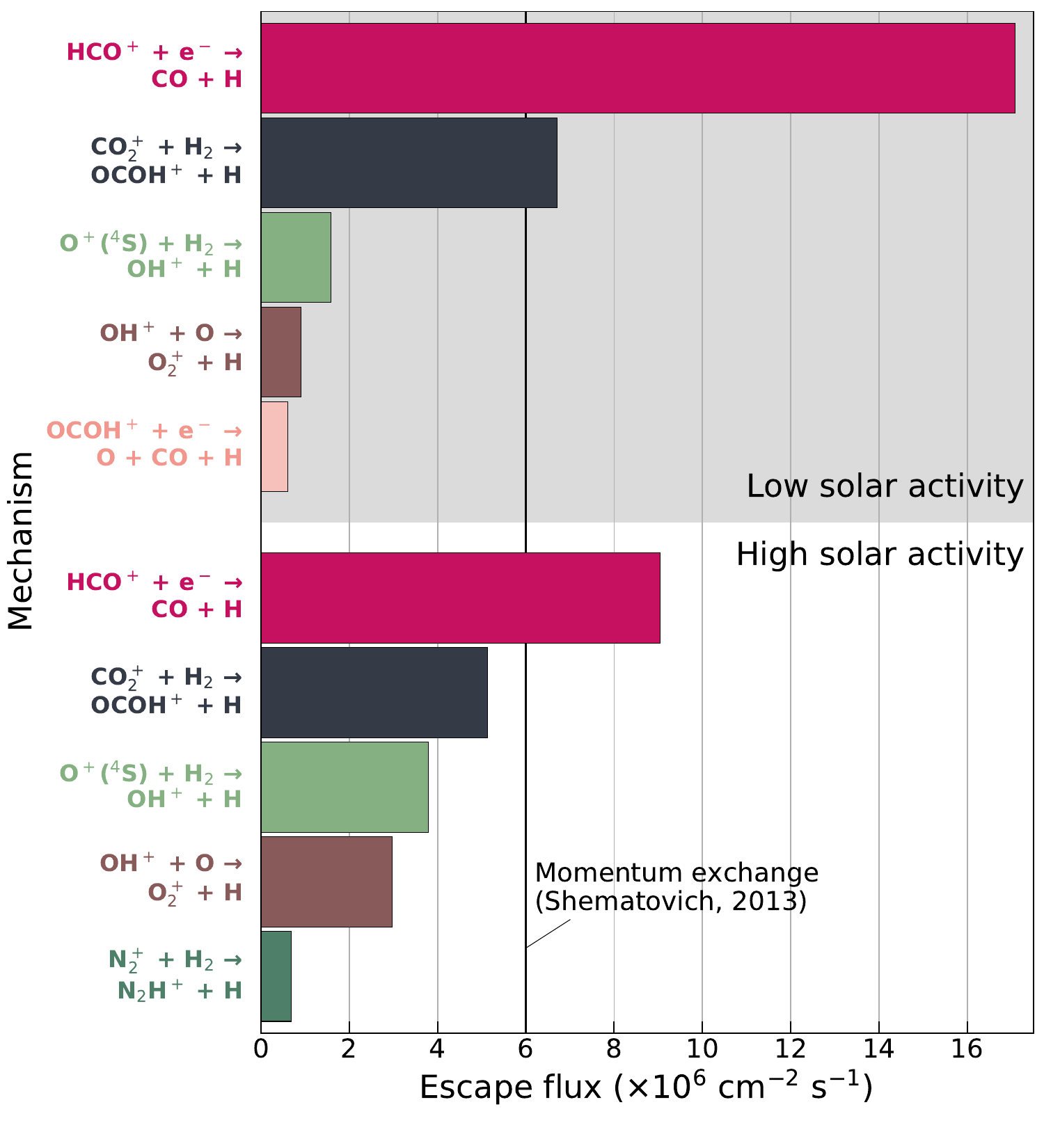}
    \caption{Escape fluxes for the most important nonthermal escape mechanisms, of those considered. Fluxes are calculated for both low and high solar activity conditions. For comparison, the black line indicates the escape flux calculated by Shematovich (2013) for hot H produced by momentum exchange with hot O.}
\end{figure} 

The profiles of the production rate of escaping hot H are shown in Figure 3 (panels (d) and (h)) for the top five most important mechanisms for each solar activity case. Also shown are the input densities ((a) and (e)) relevant for the calculation of the production rates ((b) and (f)), and the escape probabilities ((c) and (g)), as described earlier. High production rates of HCO$^+$ dissociative recombination, CO$_2^+$ + H$_2$, and OCOH$^+$ dissociative recombination below 200 km result in high production rates of escaping hot H at altitudes of 150--200 km. Below 140 km, negligible escape probabilities result in no escaping hot H produced by these mechanisms, despite high production rates. Meanwhile, the other featured mechanisms' production rate peaks are higher in the atmosphere, resulting in corresponding production rate peaks of escaping hot H at higher altitudes. The production rate of escaping hot H is tightly correlated to the reaction rate, and the escape flux is therefore closely dependent on the column production rate.

\begin{figure}[!ht]
    \centering
    \includegraphics[scale=0.33]{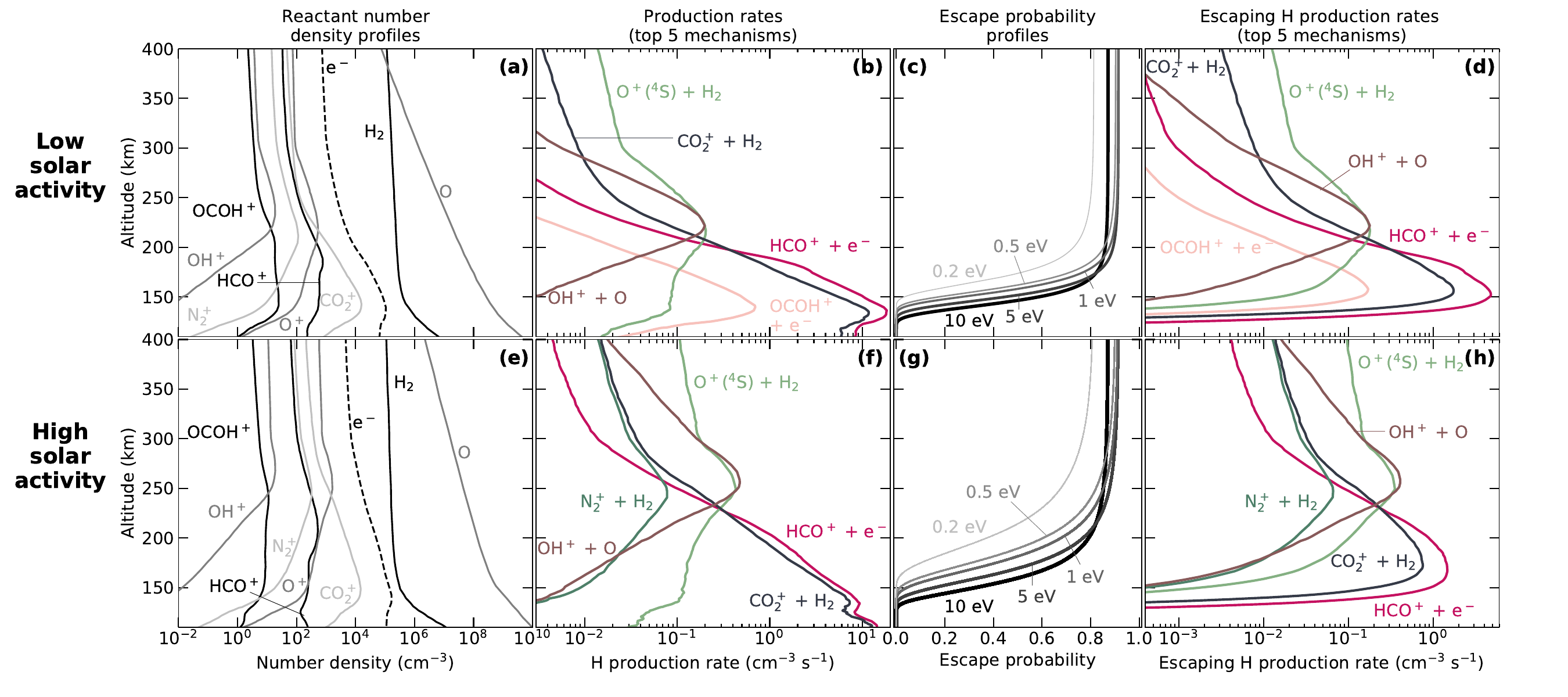}
    \caption{Input density, production rate, escape probability, and escaping hot H production rate profiles for the most important mechanisms in this study. The escaping production rate profiles, calculated using the most appropriate energy's escape probability profile, are shown for each mechanism in panels (d) and (h). The input density profiles relevant to the production of hot H via the top five selected mechanisms (from Fox, 2015), are shown in panels (a) and (e), the production rate profiles are shown in panels (b) and (f), and the escape probability profiles for five initial kinetic energies are shown in panels (c) and (g). Note that OCOH$^+$ + e$^{\text{-}}$ $\rightarrow$ O + CO + H is among the most important for low solar activity conditions (upper panels), whereas N$_2^+$ + H$_2$ $\rightarrow$ N$_2$H$^+$ + H is among the most important for high solar activity conditions (lower panels).}
\end{figure}

The left-hand panels in Figure 4 show the combined escaping H production rate for the top five reactions in this study, and the pie charts show their relative contributions to the total nonthermal escape flux. For reference, the escape flux via momentum exchange (Shematovich, 2013) is also shown. HCO$^+$ dissociative recombination produces 30--50\% of the escaping hot H. The next most important mechanisms are CO$_2^+$ + H$_2$ and momentum exchange. The production of escaping hydrogen extends over a greater range of altitudes with higher solar activity, while the peak rate decreases. The mechanisms OH$^+$ + O, O$^+$($^4$S) + H$_2$, and N$_2^+$ + H$_2$, which peak higher in the atmosphere, become more important in both absolute and relative terms under high solar activity when the thermosphere expands. In contrast, the importance of HCO$^+$ dissociative recombination, CO$_2^+$ + H$_2$, and OCOH$^+$ dissociative recombination decreases with increasing solar activity, as these mechanisms peak lower in the atmosphere. This effect is enhanced by production rates for HCO$^+$ dissociative recombination and CO$_2^+$ + H$_2$ peaking at lower altitudes under high solar activity conditions, where escape probabilities are close to zero. Increased production rates at higher altitudes for these mechanisms under higher solar activity (Figure 3b and 3f) are insufficient to counteract this.

\begin{figure}[!ht]
    \centering
    \includegraphics[scale=0.55]{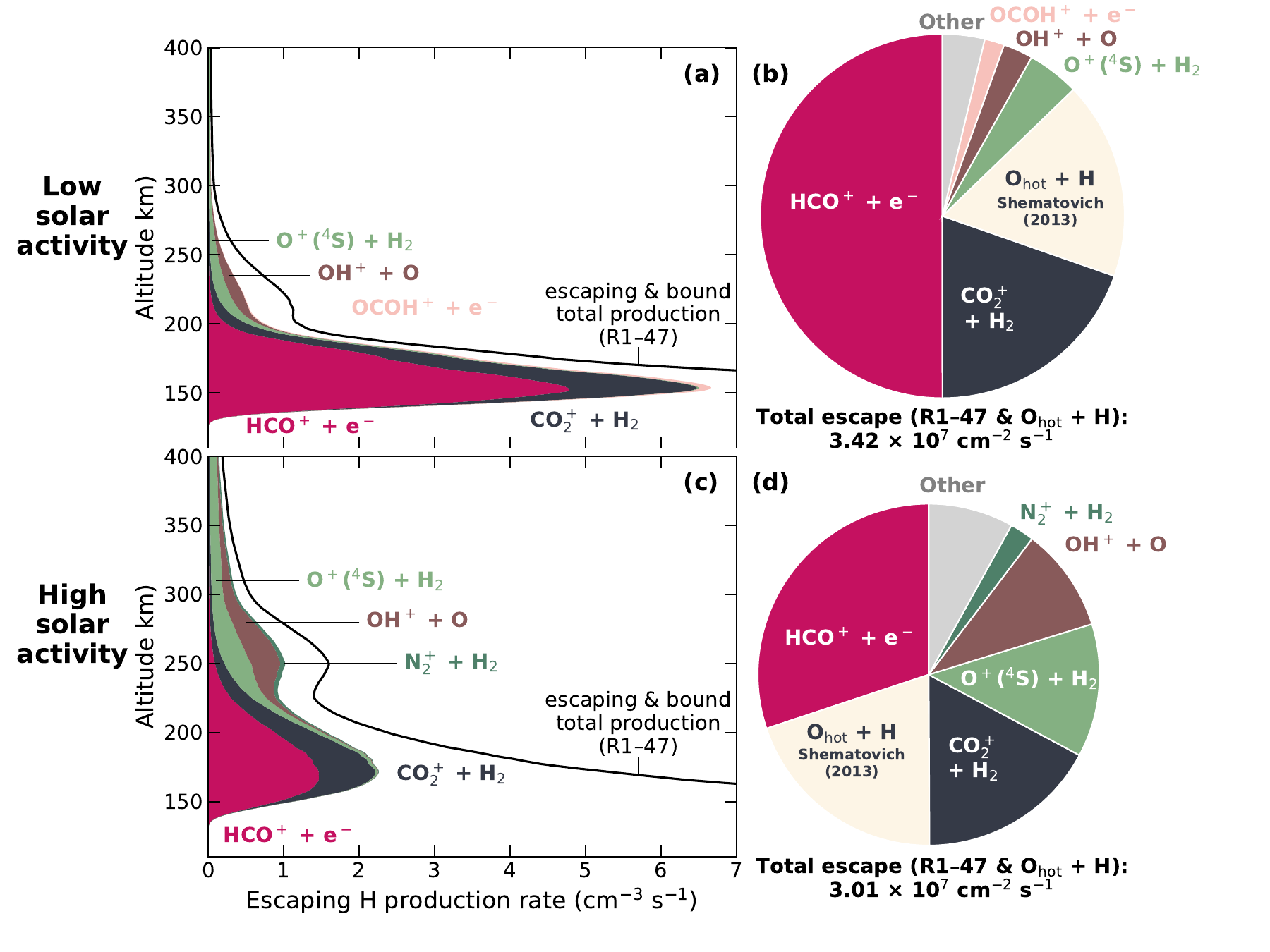}
    \begin{tabular}{@{}llll}
    \hline
      & \multicolumn{1}{l}{Thermal H escape} & \multicolumn{1}{l}{Hot H escape (47 mechanisms} & \multicolumn{1}{l}{Fraction of} \\
      & \multicolumn{1}{l}{(Fox, 2015)} &  \multicolumn{1}{l}{\& momentum exchange)} & \multicolumn{1}{l}{thermal escape} \\
      \hline
      Low solar activity & 8.7 \texttimes 10$^7$ cm$^{-2}$ s$^{-1}$ & 3.42 \texttimes 10$^7$ cm$^{-2}$ s$^{-1}$ & 39\%\\
      High solar activity & 1.1 \texttimes 10$^8$ cm$^{-2}$ s$^{-1}$ & 3.01 \texttimes 10$^7$ cm$^{-2}$ s$^{-1}$ & 27\%\\
      \hline
    \end{tabular}
    \caption{Contributions of the top five mechanisms in this study to escaping nonthermal hydrogen. Panels (a) and (c) show the cumulative escaping H production rate for the top five mechanisms for low (upper panel) and high (lower panel) solar activity. The production rate of hot hydrogen (both bound and escaping) via all 47 mechanisms considered is shown by the black line. Note the linear scale for the horizontal axis, in contrast to the logarithmic scale of Figure 3d and h. Pie charts show contributions of mechanisms towards the total nonthermal escape flux. The escape flux from momentum exchange with hot oxygen from Shematovich (2013) is included and assumed to be the same for low (upper panel) and high (lower panel) solar activity. Pie chart areas are scaled to the relative magnitudes of the total nonthermal escape flux. The table shows the total nonthermal escape flux from all 47 mechanisms and momentum exchange, compared to predicted thermal escape from Fox (2015).}
\end{figure} 

The total nonthermal fluxes from these mechanisms, including momentum exchange \cite{Shematovich2013}, are 3.42 \texttimes 10$^7$ cm$^{-2}$ s$^{-1}$ and 3.01 \texttimes 10$^7$ cm$^{-2}$ s$^{-1}$ for low and high solar activity, respectively. This equates to 39\% and 27\% of the thermal escape flux predicted by Fox's (2015) photochemical model, from which our input profiles come (8.7 \texttimes 10$^7$ cm$^{-2}$ s$^{-1}$ and 1.1 \texttimes 10$^8$ cm$^{-2}$ s$^{-1}$ for low and high solar activity, respectively; Figure 4). In both solar activity cases, more than 94\% of the escape comes from the top six mechanisms (including momentum exchange). Escape fluxes are lower for high solar activity conditions because the scale height of the background atmosphere is larger; increased background densities inhibit escape through increased frequency of collision, despite higher production rates for many mechanisms.

\section{Discussion}
\subsection{Dominant mechanisms}
Our results show that, despite its production rate peak low in the atmosphere (Figure 3b and f), the large total production from HCO$^+$ dissociative recombination makes it the dominant photochemical source of escaping hot H. We also estimate that CO$_2^+$ + H$_2$ $\rightarrow$ OCOH$^+$ + H is of similar importance as O--H momentum exchange (not included in this work). The reaction CO$_2^+$ + H$_2$ was included in the model of \citeA{Lichtenegger2006} to predict hot H densities at Mars and shown to be an important source for escaping hot H \cite{Krasnopolsky2019}. The mechanism O$^+$($^4$S) + H$_2$ was also included in the model of Lichtenegger et al. (2006) and considered a significant source of escaping H at Venus \cite{Kumar1978,Kumar1981}. The branch of OCOH$^+$ dissociative recombination to CO$_2$ + H (R14) was included in Lichtenegger et al.'s (2006) model, but not the branch to O + CO + H (R5), which we find more important here. Otherwise, escape from the photochemical mechanisms we highlight as most important in this study has not previously been calculated at Mars. Charge exchange of H$^+$ with neutral O or H, which were considered by Nagy et al. (1990) and Lichtenegger et al. (2006) and included in several studies for Venus hydrogen escape (Cravens et al., 1980; Hodges, 1999; Hodges \& Tinsley, 1981, 1982, 1986; Kumar \& Hunten, 1974; McElroy et al., 1982; Sze \& McElroy, 1975) are not within the top five most important.
\nocite{Cravens1980,Hodges1999,Hodges1981,Hodges1982,Hodges1986,Kumar1974,McElroy1982,Sze1975}

\renewcommand{\thetable}{2}
\begin{sidewaystable}
\caption{Table summarizing the five most important mechanisms contributing to nonthermal hydrogen escape, of those considered in this study. The excess energy of each mechanism (Column 3), the initial energy of the escape probability profile chosen (Column 4), the altitude
at which the production rate profiles peak, and the integrated column production rate are shown. The escape flux computed using the escape probability profile indicated is shown, along with the escape efficiency (escape flux divided by total column production rate).}
\centering
\begin{tabular}{@{}p{5mm}p{48mm}llllll}
\hline

    &    &   & Assumed initial energy &     & Total column  & &  \\
    &    &  Actual excess         & for escape probability            &  Peak production   & production & Escape flux & Escape \\
\#    & Source       &  energy [eV] &  profiles [eV] &altitude [km]& [10$^{5}$ cm$^{-2}$ s$^{-1}$] & [10$^{5}$ cm$^{-2}$ s$^{-1}$] & efficiency [\%] \\
\hline
\multicolumn{2}{l}{\textbf{Low solar activity}} &&&&&\\
\hline
R1 & HCO$^+$ + e$^{\text{-}}$ $\rightarrow$ CO + H                 & 7.31  & 5          & 137 & 789 & 171    & 21.7 \\
R2 & CO$_2^+$ + H$_2$ $\rightarrow$ OCOH$^+$ + H            & 1.35 & 1          & 135 & 496 & 67.1   & 13.5 \\ 
R3 & O$^+$($^4$S) + H$_2$ $\rightarrow$ OH$^+$ + H                 & 0.54 & 0.5        & 218 & 21.4 & 15.9	& 74.0 \\ 
R4 & OH$^+$ + O $\rightarrow$ O$_2^+$ + H                 & 1.65 & 1          & 222 & 10.1 & 9.00	& 89.0 \\
R5 & OCOH$^+$ + e$^{\text{-}}$ $\rightarrow$ O + CO + H  & 2.42 & 1 & 140  & 23.9 & 6.01	& 25.1  \\
 \hline
 \multicolumn{2}{l}{\textbf{High solar activity}} &&&&&\\
 \hline

R1 & HCO$^+$ + e$^{\text{-}}$ $\rightarrow$ CO + H        & 7.31 & 5    & 110 & 656 & 90.5 & 13.8\\
R2 & CO$_2^+$ + H$_2$ $\rightarrow$ OCOH$^+$ + H   & 1.35 & 1    & 112 & 530 & 51.4 & 9.70\\
R3 & O$^+$($^4$S) + H$_2$ $\rightarrow$ OH$^+$ + H        & 0.54 & 0.5  & 249 & 52.8 & 37.9 & 71.8\\
R4 & OH$^+$ + O $\rightarrow$ O$_2^+$ + H        & 1.65 & 1    & 256 & 34.9 & 29.6 & 85.0\\
R6 & N$_2^+$ + H$_2$ $\rightarrow$ N$_2$H$^+$ + H    & 2.61 & 1    & 247 & 8.70 & 6.91 & 79.4\\
\hline
 \end{tabular}
\end{sidewaystable}

We consider the influence of four key factors on the escape fluxes: (1) initial kinetic energy of the H atoms; (2) peak production altitude; (3) total column production rate; and (4) solar activity. Table 2 compares metrics (1)--(3) for the top five mechanisms in each solar activity case. Unsurprisingly, the escape fluxes increase with total column production, as the amount of hot hydrogen available to escape increases. The two mechanisms producing the most escaping hydrogen are those with the largest production rates, despite having peak production altitudes below 150 km. More broadly, the top seven mechanisms for escape are the top seven for production (Table S1). Peak production at higher altitudes increases a given mechanism's escape efficiency (escape flux divided by column production), as the particles will experience fewer collisions before they can escape. For mechanisms with similar peak production altitudes (e.g., HCO$^+$ dissociative recombination and CO$_2^+$ + H$_2$ (R1 and R2), or O$^+$($^4$S) + H$_2$, OH$^+$ + O, and N$^+_2$ + H$_2$ (R3, R4, and R6)), the escape efficiency decreases with decreasing initial energy, illustrating the secondary influence of energy on eventual escape. For most mechanisms, escape flux increases with reaction excess energy (Figure S4; Table S1). This effect is most pronounced in  HCO$^+$ dissociative recombination and CO$_2^+$ + H$_2$, which have high production lower in the atmosphere, where escape probability increases with energy. Conversely, the initial energy of the particles has little effect on the fluxes from O$^+$($^4$S) + H$_2$, OH$^+$ + O, and N$_2^+$ + H$_2$, as their production rates peak higher in the atmosphere where the escape probabilities are fairly similar.

The mechanisms O$^+$($^4$S) + H$_2$ , OH$^+$ + O, and N$_2^+$ + H$_2$ become more important with increasing solar activity. Increased densities of O$^+$($^4$S), OH$^+$, O, and N$_2^+$ with solar activity result in higher production rates throughout the atmosphere, especially at their peak, and the peak production altitude is relatively high, where escape probabilities are larger. This means that these processes are likely to have been more important earlier in solar system history when solar fluxes were larger \cite{Ribas2005,Tu2015}. Conversely, HCO$^+$ dissociative recombination, CO$_2^+$ + H$_2$, and OCOH$^+$ dissociative recombination become less important with increasing solar activity because the production rate at the peak decreases, even though production at higher altitudes, where escape probabilities are higher, increases.

From our results we can speculate about how our findings might apply to other terrestrial worlds. We have shown that there are myriad hot H-producing mechanisms that can contribute to nonthermal escape at Mars, and several that have not previously been considered are among the most important. Nonthermal escape is dominant for neutral H loss at Venus, where a cold thermosphere, similar to Mars', combined with large gravitational acceleration inhibits thermal escape (Donahue \& Hartle, 1992; Hodges \& Tinsley, 1981, 1986; Lammer et al., 2006b). Several photochemical mechanisms (e.g., R2, R3, R14, R27, and momentum exchange with hot oxygen) have been included in calculations for hot H at Venus (e.g., Cravens et al., 1980; Kumar \& Hunten, 1974; Lammer et al., 2006b; McElroy et al., 1982; Sze \& McElroy, 1975), and charge exchange between H$^+$ and O (R8) or H (R11) atoms are regarded as among the most important mechanisms for escape (Donahue \& Hartle, 1992; Hodges, 1999; Hodges \& Tinsley, 1981, 1986; Lammer et al., 2006b, Rodriguez et al., 1984). However, our results indicate that charge exchange with O (R8) is eighth most important under low solar activity conditions and tenth under high solar activity conditions at Mars, contributing 0.4\% and 0.8\% of our total nonthermal escape flux. Resonant charge exchange (R11) is only the 11$^{\textrm{th}}$ (24$^{\textrm{th}}$) most important mechanism for low (high) solar activity conditions, contributing 0.3\% (0.02\%) of our total nonthermal escape flux. The composition, temperature, and structure of the atmosphere of Venus are not the same as at Mars, but the same ionospheric reactions are likely to be important at both planets \cite{Lichtenegger2006}. We therefore suggest that some of the other dominant nonthermal mechanisms in this study may be more important at Venus than previously thought, especially HCO$^+$ dissociative recombination, for which escape has never been computed at Venus. They may also play a key role in CO$_2$-rich atmospheres around Venus-sized exoplanets, where nonthermal escape is likely to dominate.
\nocite{Cravens1980,Donahue1992,Hodges1999,Hodges1981,Hodges1982,Hodges1986,Kumar1974,Kumar1978,Lammer2006,McElroy1982,Rodriguez1984,Sze1975}
\nocite{Donahue1992,Hodges1981,Hodges1986,Lammer2006}
\nocite{Donahue1992,Hodges1999,Hodges1981,Hodges1986,Lammer2006,Rodriguez1984}

\subsection{Implications for total nonthermal escape}
Our accumulated nonthermal escape fluxes equate to 27--39\% of the thermal H flux predicted by Fox (2015), the study from which we obtain our input density and temperature profiles. However, the magnitude of the thermal escape flux is uncertain and variable on seasonal timescales. If the assumed average thermal escape flux is higher (e.g., 1.4 \texttimes 10$^8$ cm$^{-2}$ s$^{-1}$ \cite{Chaufray2008}; 1.8 \texttimes 10$^8$ cm$^{-2}$ s$^{-1}$ \cite{Anderson&Hord1971}; 2.4 \texttimes 10$^8$ cm$^{-2}$ s$^{-1}$ \cite{Nair1994}), the significance of the nonthermal flux decreases to 13--14\%. On the other hand, it is 60--68\% of the lowest reported seasonal escape flux, which occurs just after aphelion \cite{Bhattacharyya2015,Chaffin2014,Chaffin2015,Halekas2017}, although such seasonal conditions might simultaneously affect the ionospheric chemistry, resulting in a short-term reduction of photochemical escape fluxes below the year-round mean we calculate.

Lyman alpha brightnesses predicted from the hot H densities resulting from HCO$^+$ dissociative recombination (up to 12 Rayleighs; Gregory et al., 2023a) are far too small relative to the total hydrogen brightness (several kilo-Rayleighs; e.g., Anderson \& Hord, 1971; Bhattacharyya et al., 2015; Chaffin et al., 2014, 2018; Clarke et al., 2014) and interplanetary hydrogen brightness (200--800 R; Bertaux \& Blamont, 1971; Qu\'emerais et al., 2013; Thomas \& Krassa, 1971) for the hot H to be distinguished close to the planet. Although densities do not necessarily scale linearly with escape \cite<e.g.,>{Bhattacharyya2015, Chaffin2018}, our discovery that 30--50\% of the hot H escape flux comes from HCO$^+$ dissociative recombination suggests that the hot H produced by all the nonthermal H-producing mechanisms considered in this study would be no more than three times brighter than that predicted for H from HCO$^+$ dissociative recombination alone. This would still be too low for existing remote sensing instruments to definitively detect. One thing to note, however, is that the criteria for the Monte Carlo model are strict, such that all hot H below escape energy (i.e., hot H bound to the planet) is discounted by the model, because of the difficulty in defining the difference between a bound hot particle and an atom in the high-energy tail of a thermal distribution. A calculation of the average escape efficiencies, weighted by mechanism production rate, indicates that 20\% of the hot H particles produced by the 47 mechanisms escape, so densities and brightnesses could be up to five times larger than our estimates, if all H below escape energy was included (i.e., $\sim$180 R). The brightness contributed by the top two reactions could be up to ten times more than calculated by the model, since their low peak production rate altitudes result in low escape efficiencies (Table 2). The contributed brightnesses could be larger still if the produced (bound and escaping) hot H can effectively transfer momentum to thermal H to produce secondary hot atoms, though this is unlikely to be important in the collisional region of the atmosphere where CO$_2$, N$_2$, and O are much more abundant as collision partners. We additionally note that, if bound hot particles were included, the density (and brightness) contributions would not scale uniformly with altitude, with bound hydrogen likely to remain closer to the planet. Given the importance of predicting hot H densities, and given these complications, future work should focus on how best to account for the bound particles.
\nocite{Bertaux1971,Quemerais2013,Thomas1971}

While scattering from the hot H may not be distinguishable from the background Lyman alpha radiation, the particles may be fast enough for a spectral shift to be detectable with sufficiently high resolution instrumentation. For examples of the velocity distributions of the hot hydrogen produced by the most important mechanism, HCO$^+$ dissociative recombination, see Gregory et al. (2023a).

We predominantly focused this analysis on hot H-producing reactions included in the photochemical model of Fox (2015), but charge exchange with penetrating protons from the solar wind could be a significant source of hot neutrals, including secondary hot particles, that we have not included \cite{Halekas2015,Jakosky2018,Krasnopolsky2010,Shematovich2021}. Additionally, other Martian atmosphere modeling studies feature other H-producing reactions that we have not incorporated. For example, \citeA{Krasnopolsky2019} and \citeA{Stone2020} include reactions subsequent to the destruction of upper atmospheric water vapor via photodissociation and other chemical pathways. These mechanisms produce more hot H during southern summer conditions, when mesospheric and thermospheric water concentrations are elevated as a result of increased temperatures and dust events \cite{Aoki2019,Chaffin2021,Federova2018,Federova2020,Neary2020,Shaposhnikov2019}, which cause increased H escape due to elevated H production \cite{Chaffin2017,Heavens2018,Stone2020}. \citeA{Krasnopolsky2019} included five H-producing exothermic reactions involving H$_2$O$^+$ and H$_3$O$^+$, with excess energies of up to 6.42 eV, peak production altitudes of up to 158 km, and total column production rates of up to 3--4 \texttimes 10$^{7}$ cm$^{-2}$ s$^{-1}$. The most productive of these mechanisms, H$_3$O$^+$ + e$^{\text{-}}$ $\rightarrow$ OH + H + H, produces H atoms with energies of up to 1.31 eV at around two thirds the rate of HCO$^+$ dissociative recombination. However, its low peak production altitude (100--110 km) and the fast decrease in production rate by over two orders of magnitude between 110 km and 150 km mean that it is unlikely to be an important mechanism for hot H escape, even during seasonal conditions conducive to high water concentrations.

\subsection{Effect of escape probability profiles and energies}
Due to the large number of mechanisms we investigate, modeling of the exact kinetic energies imparted to the hydrogen atoms for each reaction, dependent on energy distribution to the products' electronic, vibrational, and rotational states, is outside the scope of this study. It is therefore worth discussing the following: (1) how accurate is the escape estimate approach utilized here when hydrogen atoms in the real Mars system are produced with a slightly different velocity, and (2) how reliable are the escape probability profiles?

We have shown (e.g. Figure S4; Table S1) that (1) for energies between 0.2 eV and 5 eV, escape fluxes increase with particle energy for a given mechanism and (2) particle energy has the greatest effect on escape fluxes when production rates peak lower in the atmosphere. We can divide the mechanisms somewhat arbitrarily into those whose production rates peak at high altitudes and those whose production rates peak at low altitudes, with the dividing altitude chosen as the calculated exobase for 5 eV H atoms ($\sim$177 km and $\sim$204 km for low and high solar activity, respectively; see Section 3.2). Of the 47 mechanisms, 5 (11) have their production rate peaks below the exobase at low (high) solar activity, so $\sim$10--25\% of the processes are likely to show differences between fluxes when different energies are used.

For some mechanisms, a fraction of the excess energy from the reaction will be translated to elevate the velocity of the second and/or third product, as well as its electronic, vibrational, or rotational states, resulting in lower-velocity hot atoms. However, the choice of escape probability curve does not greatly affect our main results. For the most important mechanism, HCO$^+$ dissociative recombination, Gregory et al.'s (2023a) Monte Carlo modeling with careful treatment of the branching ratios to different electronic states gives an escape rate only 8--9\% lower than our estimate using the 5 eV escape probability curve alone. For the total nonthermal flux from all considered mechanisms, the choice of best probability profile has an effect of at most a factor of two.

However, our new database of escape fluxes for each mechanism for all five energies (Table S1) is a helpful new reference for accurate future escape estimates that account for branching ratios of reactions to different excited-state products. For example, HCO$^+$ dissociative recombination can produce CO in at least three different electronic states, leaving different amounts of remaining energy available for the hydrogen \cite{Rosati2007}. Instead of using the 5 eV escape probability curve alone, we can assume two or more populations of nonthermal hydrogen atoms with different energies. We can assume (1) that the branching ratio to the excited-state CO molecule CO($a^3\Pi_r$) is 23\%, with excess energy 1.30 eV, (2) that the vibrational states for this branch are 0.45 (v=0), 0.21 (v=1), 0.13 (v=2), 0.1 (v=3), 0.1 (v=4), and 0.01 (v $>$ 4) \cite{Rosati2007}, and (3) that the rest of the CO is produced in the ground state electronic and vibrational levels, with a maximum excess energy of 7.31 eV. Using our escape probability curves, we approximate the escape flux as 77\% of the 5 eV escape flux for HCO$^+$ dissociative recombination (Table S1) added to 15\% of the 1 eV escape flux, 5\% of the 0.5 eV escape flux, and 2.2\% of the 0.2 eV escape flux. This results in new escape estimates of 1.56 \texttimes 10$^7$ cm$^{-2}$ s$^{-1}$ and 8.21 \texttimes 10$^6$ cm$^{-2}$ s$^{-1}$, for low and high solar activity, respectively, which match exactly (within 1.2\%) the escape rates found using the Monte Carlo model.

We now consider the reliability of the escape probability profiles used. To produce them, we (1) assumed that H collides with CO, CO$_2$, N$_2$, and O only; (2) neglected secondary hot particles, produced by energy transfer from hot H to ambient H atoms; (3) considered only elastic collisions, rather than inelastic and quenching collisions and collisions resulting in chemical reactions consuming H; and (4) used the O--H collision cross sections \cite{Zhang2009} for collisions with CO, CO$_2$, and N$_2$ as well as O, due to a lack of dedicated cross section data. Differential cross sections for four energies (0.1 eV, 1 eV, 5 eV, and 10 eV) were used to construct cumulative distribution functions which were randomly sampled to obtain scattering angles: the cross section for the energy closest to the particle energy was employed, with no interpolation.

Figure 3 shows that, for particles originating lower in the atmosphere, escape probability increases with initial kinetic energy, while this trend breaks down at higher altitudes (above 170--220 km). In order to ascertain the main factor influencing the relationship between energy and escape probability in our Monte Carlo model (i.e., number of collisions before thermalization, probability of collision, or deflection after collision), we set up an artificial experiment designed to test the albedo of the atmosphere to 0.2--10 eV H atoms. We initialized 1000 particles with identical, downwards-directed velocity vectors from the subsolar point, at an altitude of 400 km, to investigate the non-linear relationship between energy and escape probability at higher altitudes (Figure 3c, g). We varied the initial particle energy from 0.2 eV to 10 eV, ran the model ten times for each energy, and counted the number of particles that escaped. We repeated the experiment with different total and differential cross sections, where the total cross section (dependent on energy) controls the likelihood of collision, and the differential cross section (dependent on energy and scattering angle) controls the change in direction of velocity after collision. For a first test case, we used a total cross section of 3.89 \texttimes 10$^{-15}$ cm$^{2}$ for all energies (gray line on Figure S3a). For a second test case, we used the shape of the 5 eV differential cross section data for all energies (green line on Figure S3b). Note that, since our model uses the shape of the differential cross sections rather than the actual values, it does not matter that the integrated absolute differential cross sections do not equal the total cross sections for each energy, especially given that the aim of this test is only to assess the effect of the choice of cross sections. For a third test case, we combined the changes and used the adjusted total and differential cross sections together.

\nocite{Zhang2009}

\renewcommand{\thetable}{3}
\begin{table}
\caption{Results of experiment to deduce the effect of total and differential cross sections on escape probability. In all cases, 1000 particles are initialized at 400 km and given identical velocity vectors directed straight down towards the planet from the subsolar point. In the standard case, we use the energy-dependent total cross sections and differential cross sections for energies of 0.1, 1, 5, and 10 eV from Zhang et al. (2009). In a first test case, we instead hold the total cross section constant at 3.89 \texttimes 10$^{-15}$ cm$^{2}$ for all energies (Figure S3a). For a second test case, we use the 5 eV differential cross sections (Figure S3b) for all energies. For a third test case, both the total and differential cross sections are the same for all energies. The mean escape probability of ten runs is shown. The uncertainty given is the standard deviation of the mean.}
\centering
\begin{tabular}{@{}p{9mm}cccc}
\hline
Energy & \multicolumn{4}{c}{Escape probability}\\
\cline{2-5}
 (eV)      &&&\\
 & & Case 1: Total cross & Case 2: 5 eV  & Case 3: Same total  \&\\
       & & section constant  &  differential cross    &    differential (5 eV) cross   \\
       &  Standard case  &  for all energies & section for all energies &  sections for all energies\\
       \hline
0.2 & 0.511 $\pm$ 0.011 & 0.520 $\pm$ 0.012 & 0.581 $\pm$ 0.017 & 0.586 $\pm$ 0.013  \\
0.5 & 0.693 $\pm$ 0.009 & 0.710 $\pm$ 0.013 & 0.746 $\pm$ 0.014 & 0.751 $\pm$ 0.012  \\
1   & 0.726 $\pm$ 0.020 & 0.745 $\pm$ 0.014 & 0.774 $\pm$ 0.014 & 0.786 $\pm$ 0.012  \\
5   & 0.709 $\pm$ 0.015 & 0.743 $\pm$ 0.008 & 0.806 $\pm$ 0.014 & 0.835 $\pm$ 0.006  \\
10  & 0.565 $\pm$ 0.011 & 0.602 $\pm$ 0.010 & 0.812 $\pm$ 0.012 & 0.852 $\pm$ 0.014  \\
\hline\\
\end{tabular}
\end{table}

The mean and standard deviation of the mean for ten runs for each energy and case are shown in Table 3. When the total cross section is constant with energy (Case 1), the relationship between energy and escape probability varies little from the standard case, with a small increase in escape for the 5 eV and 10 eV tests. The 5 eV experiment under Case 1 differs from that in the standard case because in the latter, as particles undergo collisions, the total cross section varies with the resulting energy degradation, whereas the total cross section remains constant with decreasing energy in Case 1. There is a large change between the standard case and the cases in which the same differential cross sections were used regardless of particle energy (Cases 2 and 3). In particular, escape in Cases 2 and 3 for the 5 eV and 10 eV particles is higher than in the standard case and Case 1, and the escape of 10 eV particles is no longer lower than that for the 5 eV particles. For Cases 2 and 3, the mean escape fraction increases with initial energy, though escape for the 5 eV and 10 eV populations is indistinguishable given the uncertainties in the model. These uncertainties could be somewhat reduced by increasing the number of particles in the artificial test simulations, but an additional test (not shown here), in which we increased the number of particles by an order of magnitude for the five energies in the standard case, did not change the results and only reduced the standard deviation by a factor of two to five, while greatly increasing the run time. The results of Table 3 highlight the influence of the choice of differential cross sections on the shape of the escape probability profiles and motivate future work determining differential cross sections, for both more collision pairs (e.g. H--CO, H--N$_2$, and most importantly H--CO$_2$) and more particle energies.

\subsection{Effects of chosen inputs: densities, temperatures, and rate coefficients}
The production rate profiles strongly determine the output flux estimates and are affected by our choice of input profiles and rate constants. Firstly, we extended density profiles for 22 species by linearly extrapolating in altitude--log(density) space, in order to obtain complete density profiles between 80 and 400 km (Figure S2). This extrapolation is unlikely to affect the results significantly, as the regions in which the densities were extrapolated are those where they are lowest and cut off in the plots of Fox (2015). The two most important mechanisms, HCO$^+$ dissociative recombination and CO$_2^+$ + H$_2$, are not affected by this assumption. Calculated production rates for O$^+$($^4$S) + H$_2$, OH$^+$ + O, N$_2^+$ + H$_2$, and OCOH$^+$ dissociative recombination are only affected by extrapolation in the lowest 40 km, where the escape probabilities, and therefore the production of escaping particles, are negligible.

Secondly, we assume that production only occurs between 80 and 400 km. The lower production boundary is justified by very small escape probabilities below $\sim$130--150 km, meaning that hot H produced below this altitude does not contribute to the escaping H flux predictions. The upper production boundary equates to assuming an ionopause at 400 km. It is unlikely that any production rate profiles would peak above this altitude, because species densities are much lower. Further, all but zero (five) mechanisms for low (high) solar activity conditions produce 90\% of the total escaping hot H within the lowest 90\% of the production range (i.e., below 368 km). This indicates that, though 82--91\% of hot H produced above 400 km would escape, production rates are unlikely to be large and significantly affect total photochemical loss.

A third result of our choice of density profiles is that we implicitly assume the same solar input as Fox (2015) and ignore latitudinal variations, which produces a flux representative of the whole dayside hemisphere. Ionospheric density profiles have been shown to vary with latitude, as a result of, e.g., variations in solar zenith angle and proximity to crustal magnetic fields \cite{Fowler2022}, which may affect the spatial distribution of escape, but that is beyond the scope of this work.

We use rate coefficients that assume that ion populations can be adequately described by a Maxwellian distribution at a single temperature. However, recent observational evidence of suprathermal ions suggests that ion velocities can diverge from the Maxwellian above the exobase region, due to various plasma physics effects above the collisional atmosphere \cite{Fowler2018, Fowler2021, Hanley2022}. Our assumption may therefore affect the hot hydrogen production rates and resulting escape fluxes for mechanisms that peak above the exobase (75--90\% of the mechanisms). However, we use a rate coefficient dependent on the ion temperature for only six of these. The most important mechanisms that peak above the exobase are reactions 3, 4, and 6, but the adopted rate coefficients for these three reactions do not have an ion temperature dependency. Calculations accounting for ion heating effects would require kinetic modeling outside the scope of our current work, especially given that the form of the distribution function above the exobase remains unknown.

We also note that our adopted rate constant for resonant charge exchange between a hydrogen ion and a hydrogen atom (R11) uses a constant cross section of 6 \texttimes 10$^{-15}$ cm$^2$, with no energy dependence \cite{Rodriguez1984}. The energy dependence of the cross section for ion temperatures at the altitudes on which we focus affects the cross section by less than 15\% \cite{Banks1968}, which is much smaller than the uncertainty introduced by some of our other assumptions. Similarly, the energy dependence of the charge exchange cross section for O + H$^+$ (R8) is very small for the energies relevant to the Martian ionosphere \cite{Stancil1999}. Our low escape flux results for the two charge exchange reactions mean that these assumptions affect our total flux estimates negligibly.

To address all these points, we stress that this study demonstrates a simple method of estimating escape fluxes from different nonthermal mechanisms. This method can be applied to any photochemical reaction using any production rate profile derived from any chosen input densities, temperatures, and rate coefficients. New background species densities can also be utilized, since the escape probability is dependent on the overhead column density, rather than on altitude. It would be straightforward in future to apply this method to more mechanisms and to use new density and temperature profiles as they are published, especially from the growing wealth of spacecraft observations. For example, while we choose to use model-computed density and temperature profiles from Fox (2015), because self-consistent temperature and density profiles for two solar activity conditions are available, we could use ion temperature profiles derived from observations by the Suprathermal and Thermal Ion Composition (STATIC) instrument on MAVEN \cite{Hanley2021,Hanley2022} and density profiles from other models \cite<e.g.,>{Cangi2020,Krasnopolsky2019,Matta2013} or spacecraft measurements \cite<e.g.,>{Fowler2022,Lillis2017}. It is difficult to speculate in detail about how using different input profiles might affect our escape fluxes, but we expect that escape fluxes would vary to the same extent as the profiles deviate from those of Fox (2015). Our method also provides the opportunity to estimate variation of escape via different mechanisms on diurnal, seasonal, and longer-term timescales, as well as various spatial scales. To this end, the escape probabilities have already been incorporated into a full photochemical model, which can be run on demand to calculate H escape rates for different atmospheric conditions (Cangi et al., 2023, this issue).
\nocite{Cangi2023}

\subsection{Future work}
There are several parameter values that, if known better, would improve our escape flux estimates. Of the highest priority are elastic collision cross sections (total and differential) of H with CO, CO$_2$, and N$_2$, which could be determined through either laboratory or theoretical techniques. Our test described in Table 3 shows that the differential cross sections have a particularly strong control on the relationship between particle energy and escape and are therefore particularly critical.

Secondly, since escape flux is closely correlated to production rate, further model or observational constraints on density and temperature profiles would improve calculations. Further, many rate coefficients remain uncertain: for example, the 50\% uncertainty in the rate coefficient for HCO$^+$ dissociative recombination \cite{FonsecaDosSantos2014,Fox2015,Korolov2009} would affect the escape rate by a factor of two.

Thirdly, since the escape probabilities are somewhat dependent on the kinetic energy of the hot H atom produced, a better knowledge of branching ratios to products in different electronic, vibrational, or rotational states would be helpful for detailed Monte Carlo modeling of the most important processes. For example, while branching ratios to different vibrational levels of the excited-state CO produced by HCO$^+$ dissociative recombination have been published \cite{Rosati2007}, the branching ratios to the second excited state of CO, the vibrational levels of the ground state CO, and all rotational levels of the CO remain unknown. Monte Carlo simulations for the top mechanisms, incorporating the details of energy distribution to hot H for each, will give us the most comprehensive picture to date of how and why hot H is escaping at Mars, as well as allowing the prediction of densities and velocity distributions to better understand the resulting hot hydrogen population.

\section{Conclusions}
We have presented a comprehensive study of more sources of escaping nonthermal hydrogen than have been evaluated before, showing that in total, photochemical mechanisms can contribute to escape of up to 39\% the magnitude of the thermal escape flux. We show that HCO$^+$ dissociative recombination, only recently explored as a source of escaping hot hydrogen, is the dominant photochemical mechanism for hydrogen loss at Mars, constituting a third to half of the total photochemical escape flux. Despite a low peak production altitude, the high integrated column production from this reaction results in significant escape. It is likely that it is important at Venus, where charge exchange reactions between hydrogen ions and atoms, of low importance here, are thought to be dominant (Hodges \& Tinsley, 1981, 1986; Lammer et al., 2006b). Our method for estimating escape fluxes is valuable for application to any chosen input production rate profiles, including temporally- and spatially-varying values. We present a step forward in understanding of the key mechanisms behind hot hydrogen escape and their importance for Martian desiccation, as well as their variation over solar system history, providing a key insight into how the atmosphere may have evolved.

\nocite{Lammer2006,McElroy1982,Rodriguez1984}
\nocite{Hodges1981,Hodges1986,Lammer2006}

\section{Open Research}
\noindent The model code is available in Gregory et al. (2023b).

\nocite{code_and_data2023b}
\nocite{Gregory2023}

\acknowledgments
This work was supported by NASA Solar System Workings grant number 80NSSC19K0164. E. Cangi is supported by NASA’s FINESST Program (Grant \#80NSSC22K1326).

\bibliography{bibliography}

\end{document}